\documentclass[aps,pra,reprint,floatfix]{revtex4-1}

\usepackage{amsmath,amsfonts,amssymb}
\usepackage{graphicx}
\usepackage{color}
\usepackage{ulem}
\usepackage{float}
\usepackage{booktabs}

\def\beq{\begin{equation}}
\def\eeq{\end{equation}}
\def\beqn{\begin{eqnarray}}
\def\eeqn{\end{eqnarray}}

\begin{document}

\title{Ultrafast intermolecular energy transfer from vibrations to electronic motion}

\author{Lorenz S. Cederbaum}
\affiliation{Theoretische Chemie, Physikalisch--Chemisches Institut, Heidelberg University, 
Im Neuenheimer Feld 229, D-69120 Heidelberg, Germany}
\date{\today}

\begin{abstract}
	
It is discussed how vibrationally excited molecules in their electronic ground state can transfer their vibrational energy to the electronic motion of neighbors and ionize them. Based on explicit examples of vibrationally excited molecules and anionic neighbors, it is demonstrated that the transfer can be extremely efficient at intermolecular distances much beyond distances at which the molecule and its neighbor can form a bond.

\end{abstract}

\maketitle


Intermolecular energy transfer processes between an excited molecule and its neighbors are ubiquitous in nature and very widely studied. If the energy transferred is {\it electronic} and between bound electronic states of the molecule and one of its molecular neighbors, the energy transfer process is referred to as Foerster (or fluorescence) resonance energy transfer (FRET) \cite{Foerster}. Prominent examples are exciton transfer in semiconductors \cite{exciton_Fink}, and the first step of photosynthesis which involves the energy transfer from antenna complexes to reaction centers \cite{Renger_May_Kuehn,Scholes_Fleming}. As energy is conserved, FRET is only possible if nuclear motion takes place and this leads to a time scale of picoseconds or longer \cite{Renger_May_Kuehn,Scholes_Fleming}. 

Another, highly efficient {\it electronic} energy transfer mechanism, called interatomic (or intermolecular) Coulombic decay (ICD), becomes operative once the excess energy suffices to ionize the neighbor \cite{Giant}. The transferred energy ionizes the neighbor and hence energy conservation is fulfilled without the need for nuclear motion. Consequently, the excited species as well as the neighbors can be atoms or molecules and the time scale involved is in the femtosecond regime \cite{Averbukh_Review,Hergenhahn_Review,Jahnke_Review}. Applications of ICD range from ICD in extreme systems like the He dimer which is the most weakly bound system known with an average distance of 52 {{\AA}} between the atoms \cite{Nico_He_Dimer,Exp_He_Dimer}, ICD after Auger and resonant Auger processes \cite{After_Auger_Theory,After_Auger_Exp,Resonant_Auger_Theory,Resonant_Auger_Exp1,Resonant_Auger_Exp2,X_Ray_Induced}, to ICD in quantum dots and quantum wells \cite{ICD_QD_Nimrod,ICD_QD_Annika}. 

Although less investigated, intermolecular {\it vibrational} energy transfer between weakly bound molecules, i.e., at internuclear distances at which they do not have a chemical bond, is also of relevance. Here, the long-range coupling between the molecules is determined, similarly as in FRET, by the transition-dipole transition-dipole interaction, but now for the involved vibrational transitions and not for the electronic transitions, see, e.g., \cite{RVET_Water_Woutersen,VET_condensed_Chen,non_Resonant_VET}. Most of the studies were done for describing {\it resonant} vibrational energy transfer in the condensed phase. Very recently, it has been noticed that if the lifetime of the vibrationally excited molecule is much longer than that of its neighbor, efficient {\it non-resonant} vibrational energy transfer can take place \cite{non_Resonant_VET}.

In this work we would like to investigate the possibility of intermolecular vibrational energy transfer to electronic motion. Energy transfer of all kinds is of central importance for chemical reactivity and has been widely studied both experimentally and theoretically over many years including the transfer between the two kinds of energies, vibrational and electronic. The studies of the latter are, however, carried out in the framework of collisions where the collision complex formed and/or nonadiabatic coupling give rise to the transfer \cite{Flynn_1,Hippler_Troe,Flynn_2,Chandler}. Here, we concentrate on intermolecular vibrational energy transfer to electronic motion in weakly bound molecules, i.e., at internuclear distances at which they do not have a chemical bond and nonadiabatic coupling is negligible. We shall see that the transfer can be highly efficient.


Consider a molecule A in an excited vibrational state $\nu_i$ of its electronic ground state $\phi_0^{A}$ and a neighboring molecule B in a vibrational state $\nu_i'$ and electronic state $\phi_i^{B}$. Note that as we investigate vibrational to electronic energy transfer, the neighbor B can also be an atom in an electronic state $\phi_i^{B}$. The relaxation process we discuss is as follows: Molecule A relaxes from $\nu_i$ to a vibrational level $\nu_f$ of lower energy, and the excess energy is utilized to ionize the neighbor B. For convenience we choose the energy of the vibrational ground state of our system A to be the zero of the energy scale and characterize the vibrational states by their frequency, i.e., the energy of a state $\nu_i$ is $h\nu_i$. Of course, the excess energy $h\nu \equiv h\nu_i - h\nu_f$ must be larger than the energy required to remove an electron from the neighbor B. 

The rate of the process, or more precisely the decay width, is determined by the golden rule 

\begin{align}
\label{eq::1}
\Gamma = 2\pi \sum_{f} \lvert \langle {\Psi_i}| H | {\Psi_f}\rangle   \rvert^2,
\end{align}
where $H$ is the full Hamiltonian of A and B and their interaction. The wavefunctions ${\Psi_i}$ and ${\Psi_f}$ describe as usual the initial and final states of the process in the absence of the interaction between A and B. In the Born-Oppenheimer approximation, the initial state is given by the product ${\Psi_i} = \nu_i \phi_0^{A} \nu_i' \phi_i^{B}$ and the final state by ${\Psi_f} = \nu_f \phi_0^{A} \nu_f' \phi_f^{B}$. The $\nu_f'$ stands for the vibrational state of the ion B (if B is a molecule) and the electronic final state of B describes the ion B together with an electron in the continuum and is chosen to be energy normalized. The sum over the final states also includes different vibrational levels of the ion and possibly different ionic electronic states. 

In general, the above expression cannot be evaluated analytically and is rather cumbersome to be computed numerically. At large distances R between A and its neighbor B, one can evaluate the leading term of $\Gamma$ analytically by expanding the Coulomb interaction between the charged particles (electrons and nuclei) of A and B. Let ${\bf{x}}_i$ and ${\bf{X}}_k$ be the electronic and nuclear coordinates of the particles of molecule A, and ${\bf{x'}}_j$ and ${\bf{X'}}_l$ those of B in some coordinate system. The coordinates defined by ${\bf{r}}_i = {\bf{x}}_i - {\bf{R}}_A$ and ${\bf{R}}_k = {\bf{X}}_k - {\bf{R}}_A$ are those of the electrons and nuclei of molecule A relative to its center of mass ${\bf{R}}_A$. Analogously,  ${\bf{r'}}_j = {\bf{x'}}_j - {\bf{R}}_B$ and ${\bf{R'}}_l = {\bf{X'}}_l - {\bf{R}}_B$ are those of B relative to its center of mass ${\bf{R}}_B$. When describing electronic processes, like, for instance, ICD, one has to expand the electron-electron repulsion (see, e.g., \cite {res_Robin} ):
\begin{align*}
\nonumber
&\sum_{i,j}\frac{1}{|{\bf {x}}_i - {\bf {x'}}_j|} = \frac{N_A N_B}{R} + \sum_{i,j}\frac{{\bf {u}} \cdot ({\bf {r}}_i - {\bf {r'}}_j) }{R^2} \\ & + \sum_{i,j}\frac{-3 ({\bf {u}} \cdot {\bf {r}}_i)({\bf {u}} \cdot {\bf {r'}}_j) + {{\bf {r}}_i} \cdot {{\bf {r'}}_j} }{R^3} + O(\frac{1}{R^4}).
\end{align*}
Here, ${\bf {u}} = ({\bf{R}}_A - {\bf{R}}_B)/R$ is a unit vector along the line segment that joins the centers of mass of A and B and $N_A$ and $N_B$ are the number of electrons in A and B. 

When describing vibrational to electronic energy transfer, one has also to expand the interaction between the nuclei of A with those of B as well as the attraction between the electrons of A with the nuclei of B and, of course, between the electrons of B and the nuclei of A. The expressions for the attraction are somewhat lengthy, but the final expression for the total interaction is simple and takes on the appearance
\begin{align}
\label{eq::2}
\nonumber &\frac{a}{R} + \frac{{\bf {u}} \cdot ({\bf {\hat{D}}}^A - {\bf {\hat{D}}}^B) }{R^2} \\& + \frac{-3 ({\bf {u}} \cdot {\bf {\hat{D}}}^A)({\bf {u}} \cdot {\bf {\hat{D}}}^B) +  {\bf {\hat{D}}}^A \cdot {\bf {\hat{D}}}^B }{R^3} + O(\frac{1}{R^4}),
\end{align}
where $a$ is a number and  ${\bf {\hat{D}}}^A$ and ${\bf {\hat{D}}}^B$ are the dipole operators of A and B 
\begin{align}
\label{eq::3}
\nonumber &{\bf {\hat{D}}}^A = -\sum_{i}{\bf {r}}_i + \sum_{k}Z_k{\bf{R}}_k \\& {\bf {\hat{D}}}^B = -\sum_{j}{\bf r}'_j + \sum_{l}Z'_l{\bf{R}}'_l ,
\end{align}
which include all charged particles of either A or B, the nuclear charges being $Z_k$ and  $Z'_l$.

Returning to Eq.(\ref{eq::1}) one immediately notices that the first two terms in Eq.(\ref{eq::2}) do not contribute to the rate because of the orthogonality of the vibrational levels in the ground electronic state of A, $\langle\nu_i|\nu_f\rangle = 0$, and of the electronic states of B, $\langle \phi_i^{B}|\phi_f^{B}\rangle = 0$. The third term provides the leading non-vanishing contribution to the rate. Inserting this term into Eq.(\ref{eq::1}) and integrating over the electronic coordinates gives rise to the permanent dipole moment ${\bf {{D}}}_0^A$ of A, which is, of course, a function of all nuclear coordinates $\{{\bf{R}}_k\}$, and to the electronic transition dipole element ${\bf {{D}}}_{i_e,f_e}^B$ of B to the continuum which is a function of $\{{\bf{R}}'_l\}$:
\begin{align}
\label{eq::4}
\nonumber &{\bf {{D}}}_0^A = \langle \phi_0^{A}|{\bf {\hat{D}}}^A |\phi_0^{A}\rangle \\& {\bf {{D}}}_{i_e,f_e}^B = \langle \phi_i^{B}|{\bf {\hat{D}}}^B |\phi_f^{B}\rangle.
\end{align} 
Integration over the nuclear coordinates now leads to the vibrational transition dipole element ${\bf D}_{i_v,f_v}^A$ in the electronic ground state of A and to the vibrational-electronic transition dipole element ${\bf D}_{i_ei_v,f_ef_v}^B$ of B which take on the appearance
\begin{align}
\label{eq::5}
\nonumber &{\bf D}_{i_v,f_v}^A = \langle \nu_i|{\bf {{D}}}_0^A |\nu_f\rangle \\&{\bf {{D}}}_{i_ei_v,f_ef_v}^B  = \langle \nu_i'|{\bf {{D}}}_{i_e,f_e}^B |\nu_f'\rangle.
\end{align} 

The resulting expression for the rate $\Gamma$ reads
\begin{align}
\label{eq::6}
\nonumber &\Gamma = \frac{2\pi}{R^6}  \sum_{f'_e,f'_v} \lvert S \rvert^2, \\& S = 3 ({\bf {u}} \cdot {\bf D}_{i_v,f_v}^A)({\bf {u}} \cdot {\bf D}_{i_ei_v,f'_ef'_v}^B) -  {\bf D}_{i_v,f_v}^A \cdot {\bf D}_{i_ei_v,f'_ef'_v}^B. 
\end{align} 
The rate obviously depends on the orientation of the two transition dipoles and the unit vector connecting the centers of masses of A and B. Choosing  ${\bf {u}}$ to point in the direction of the z-axis and ${\bf D}_{i_v,f_v}^A$ to lie in the xz-plane of the coordinate system, it is easy to average over the orientation of A and B. This leads to $\Gamma = \frac{4\pi}{3R^6}  \sum_{f'_e,f'_v} \lvert {\bf D}_{i_v,f_v}^A\rvert^2 \lvert{\bf D}_{i_ei_v,f'_ef'_v}^B \rvert^2$. Clearly, $\Gamma$ in Eq.(\ref{eq::6}) is maximal if all three vectors are parallel, and this in turn gives rise to an increase by a factor of 6 compared to the latter. 

To make connection with experimentally measurable quantities, we replace the vibrational transition dipole $\lvert{\bf D}_{i_v,f_v}^A\rvert^2$ by the respective Einstein coefficient $A_{i_v,f_v}^A$ according to \cite{ Thorne}
\begin{align*}
\nonumber
\lvert{\bf D}_{i_v,f_v}^A\rvert^2 = \frac{3\hbar c^3}{32\pi^3\nu^3}A_{i_v,f_v}^A,
\end{align*}
where {\it c} is the speed of light, and make use of the relationship \cite{Sobel'man}
\begin{align*}
\nonumber
\sigma^B_{i_v,i_e}(h\nu) = \frac{8\pi^3}{3} \frac{\nu}{c}  \sum_{f'_e,f'_v} \lvert {\bf D}_{i_ei_v,f'_ef'_v}^B \rvert^2, 
\end{align*} 
defining the total photoionization cross section of B for photons of energy $h\nu$. One readily obtains
\begin{align}
\label{eq::7}
\Gamma = \alpha \frac{3\hbar}{4\pi} \big(\frac{c}{2\pi \nu}\big)^4 \frac{A_{i_v,f_v}^A \sigma^B_{i_v,i_e}}{R^6}, 
\end{align}
where $\alpha = 1$ for randomly oriented molecules and $\alpha = 6$ for ideally oriented molecules.

Let us briefly discuss the above result. Molecule A relaxes from its initial vibrational state $\nu_i$ to its final $\nu_f$ state and the excess energy $h\nu=h\nu_i - h\nu_f $ is utilized to ionize the atomic or molecular neighbor B. The rate of this process, of course, grows the 'easier' the neighbor B can be ionized, i.e., the larger is its photoionization cross section in the initial electronic-vibrational state it has been in before the relaxation. As in the purely electronic ICD process \cite{Averbukh_PRL}, one may also view the vibrational-electronic process as ionization by a virtual photon. We suggest to call this process vibrational ICD. This virtual photon is emitted by the vibrationally excited molecule A. The Einstein coefficient $A_{i_v,f_v}^A$ is the inverse radiative lifetime $\tau_{rad}$ of the vibrational state $\nu_i$ of molecule A due to the $\nu_i \rightarrow \nu_f$ transition, $A_{i_v,f_v}^A = 1/\tau_{rad}$, and consequently, the vibrational ICD becomes faster the faster the radiative decay of the isolated molecule A is. If there are more than one final vibrational state $\nu_f$ of A which leads to the ionization of B, then the rate in Eq.(\ref{eq::7}) becomes a partial width $\Gamma_{f_v}$ and the total width of the decay of the initial state $\nu_i$ is simply the sum over all partial widths: $\Gamma = \sum_{f_v}\Gamma_{f_v}$. The lifetime of the initial state due to the vibrational ICD becomes $\tau = \hbar / \Gamma$. 

Finally, we give the vibrational ICD rate and lifetime in units appropriate for the process at hand: 
\begin{align}
\label{eq::8}
\nonumber&\Gamma[{cm^{-1}}] = \alpha \times \frac{8.103 \times 10^{14}}{(\nu[{cm^{-1}}])^4}  \frac{\big(A^A[{s^{-1}}]\big) \big(\sigma^B[Mb]\big)}{(R[{\AA}])^6}, \\& \tau [s] = \frac{5.31 \times 10^{-12}}{\Gamma [cm^{-1}]}, 
\end{align}
where, for brevity, indices have been dropped.


Vibrationally excited molecules in the
electronic ground state decay rather slowly radiatively. 
Their radiative lifetime is typically in the range of seconds
to milliseconds \cite{vib_Radzig_Smirnov,rad_lifetime_Tennyson}. This reduces the vibrational ICD rates compared to the purely electronic ICD where the radiative lifetime of the involved electronic states is typically in the ns time scale. On the other hand, vibrational energies are much smaller than the electronic energies and as $h\nu$ enters to the fourth power in the rate, there is {\it a priori} hope that vibrational ICD can be efficient. Before turning to concrete examples, vibrational ICD is schematically depicted in Fig.1.

\begin{figure}[!]
	\includegraphics[width=1.0\columnwidth,angle=0]{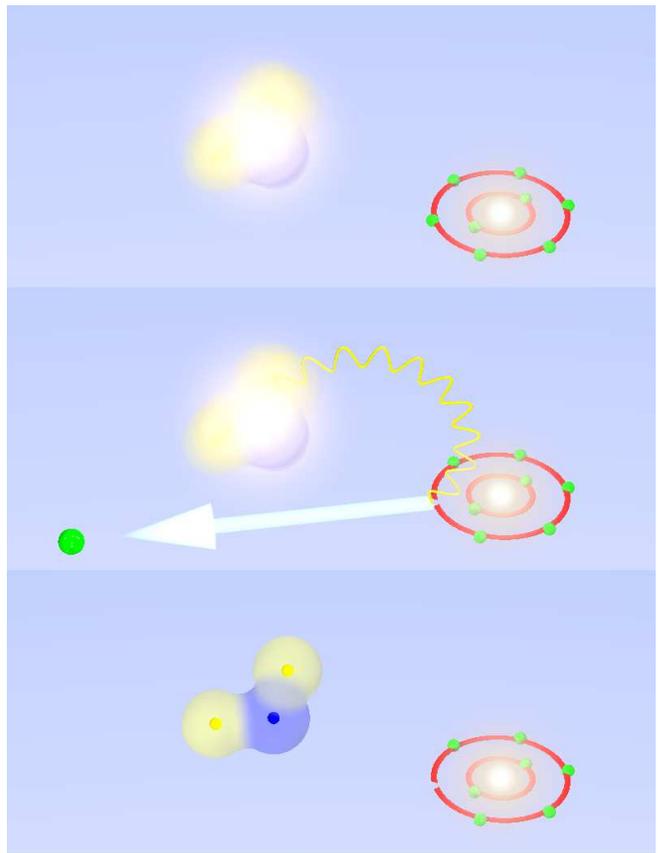}
	\caption{(Color online) Vibrational ICD. Upper panel: The left molecule is in its electronic ground state and is vibrationally excited while the neighbor is in its ground state. Middle panel: The vibrational excess energy of the left molecule is transferred to the neighbor and ionizes it. Lower panel: The molecule is now in its electronic and vibrational ground state and the neighbor possesses one electron less. At large distances between the molecule and its neighbor the rate of vibrational ICD is given in closed form in Eq.(\ref{eq::7}). Note that the equation formally also covers cases where the neighbor is in an excited state. The figure is by courtesy of Till Jahnke.} 
	\label{fig1}
\end{figure}

There is great interest in the literature in atomic and molecular negative ions and we shall use them here as a proof of principle for vibrational ICD. The importance and utility of negative ions extend well beyond the regime of gas-phase ion science, e.g., \cite{Gas_Phase_Ion_Chem}, materials, and environmental science \cite{Materials_Science_Ion}, pure chemistry \cite{Pure_Chemistry_Ion} and more \cite{Andersen_Review,EA_Schaefer}. In contrast to neutral systems where the photoelectron experiences the influence of the long-range Coulomb potential, the electron detached from a negative ion sees a short-range potential exerted by the residual neutral system. The photoionization cross section of neutral systems is non-zero and often dominant at threshold and that of negative ions (usually referred to as photodetachment cross section) depends strongly on the angular momentum of the detached electron \cite{Wigner,Hotop,Andersen_Review}. For atoms, unless the ejected electron originates from a p-orbital, the cross section is zero at threshold and then it grows as a function of photon energy, reaches a maximum and falls off. For s-wave photoelectrons, the cross section is typically large at threshold. Molecules do not posses spherical symmetry and one can expect a substantial non-zero photodetachment cross section at threshold. 

A few words on the binding of negative ions. Most, but not all atoms posses bound negative ions \cite{Andersen_Review,EA_Schaefer}. Some, even 'standard' molecules like water and benzene do not bind an electron. In clusters, such molecules may form stable anions with stability growing with the cluster size, see, e.g., \cite{Stokes_Compton_nitromethane,EA_Schaefer,Vysotskiy_Water} and references therein. There are weakly bound anions where the excess electron is held by the molecule's dipole (dipole-bound) or by electron correlation (correlation-bound) \cite{Dipole_bound,Corre_bound}. Interestingly, some biologically relevant molecules like nucleobases have dipole-bound anions which become valence-bound when microsolvated \cite{Anion_Nucleobases}. All of this makes clear that a plethora of interesting weakly bound negative ions exists. 

In the following we discuss examples of vibrationally excited typical molecules and anions as neighbors. We start with weakly bound anions, where a single vibrational quantum suffices for vibrational ICD, and progress to more strongly bound ones where more than one quantum is needed.  

The interest to study alkaline-earth-metal negative ions has been stimulated by the discovery that the closed shell Ca atom forms a stable Ca$^-$ negative ion  \cite{Pegg_Ca,Fischer_Ca}. For Ca$^-$ experimental and theoretical absolute photodetachment cross sections are available \cite{Andersen_Review,Yuan_Ca} which allow the evaluation of the vibrational ICD rate. The binding energy of Ca$^-$ is taken to be 24.55 meV \cite{EA_Ca}. The bending mode of the water molecule has a frequency $\nu$ of 1594.8 cm$^{-1}$ and an Einstein coefficient of 16 s$^{-1}$ \cite{vib_Radzig_Smirnov}. At the respective photon energy of $h\nu = 0.20$ eV, the detachment cross section amounts to 130 Mb \cite{Andersen_Review,Yuan_Ca}. Employing Eq.(\ref{eq::8}), one finds a decay width of 0.26 cm$^{-1}$ at a distance R = 1 nm. That at such a large distance the lifetime of the bending mode reduces by vibrational ICD from 62 ms to 20 ps is indeed amazing. This lifetime is even well shorter than radiative lifetimes of electronic transitions which are typically in the ns range. Choosing a partner with a lower frequency can make the effect even more dramatic. Taking HCN instead of H$_2$O, the bending frequency is 712 cm$^{-1}$, $A^A = 15$ s$^{-1}$ and $\sigma^B = 135$ Mb \cite{vib_Radzig_Smirnov,Andersen_Review,Yuan_Ca}, the lifetime of the energy transfer reduces from the radiative lifetime of 69 ms to just 0.8 ps at the distance of 1 nm. And this all is for randomly oriented molecules, i.e., $\alpha = 1$ in Eq.(\ref{eq::8}). Choosing a molecule with an even smaller vibrational frequency does not necessarily imply a further larger impact. The frequency of the 12E$_u$ mode of C$_2$F$_6$, for example, is 220 cm$^{-1}$, but its radiative lifetime is very long (78 s) \cite{vib_Radzig_Smirnov}. With $\sigma^B = 140$ Mb \cite{Andersen_Review,Yuan_Ca}, the resulting ICD lifetime is 8.4 ps at R = 1 nm. 

How to lower the frequency and still enlarge the Einstein coefficient? Nowadays, there are experimental techniques to produce high vibrational levels (overtones and combination tones) of molecules in their ground electronic state, e.g., \cite{Lehmann,Snels_Quack}. The resulting excess vibrational energy may suffice to ionize via vibrational ICD most of the negative ions available. However, the Einstein coefficient corresponding to the transition from such a high level to the ground vibrational level is typically smaller than that for a transition from a low vibrational level, see, e.g. \cite{Einstein_High_Level}, This, and the larger excess energy make the vibrational ICD much less favorable, see Eq.(\ref{eq::7}). On the other hand, Einstein coefficients of high vibrational levels for transitions to close by levels can be substantially larger than those for transitions from a low vibrational level. For instance, for the molecule NaCl the coefficient for the transition from the eighth vibrational level to the seventh is $A_{8,7}^A = 6.1$ s$^{-1}$ while that from the first to the ground state is $A_{1,0}^A = 0.9$ s$^{-1}$ \cite{Einstein_NaCl}, and for CO the situation is even more pronounced: $A_{12,11}^A = 250$ s$^{-1}$ while $A_{1,0}^A = 34$ s$^{-1}$ \cite{vib_Radzig_Smirnov}. At the same time the higher transition frequencies are reduced by anharmonicity, for NaCl from 361.6 to 345.0 cm$^{-1}$ \cite{Frequencies_NaCl}. With Ca$^-$ as a neighbor ($\sigma^B = 140$ Mb \cite{Andersen_Review,Yuan_Ca}) the relaxation time of NaCl becomes extremely fast, namely 0.88 ps and even 110 fs (!) at R = 1 nm for the $1\rightarrow0$ and $8\rightarrow7$ transition, respectively. The relaxation in the latter case is still amazingly fast, 6.7 ps, at the large distance of 2 nm. 

There is much interest in nitric oxide anion, NO$^-$, because of its important physiological role \cite{NO_Anion_Chem}. Its electron binding energy is 26 meV \cite{NO_Anion_BE} and thus similar to that of Ca$^-$. There is relevant work on the relative photodetachment cross section \cite{NO_Anion_PD} from which I could only estimate the absolute values to be about 10 times smaller than those of Ca$^-$. Consulting Eq.(\ref{eq::7}) and the above results for Ca$^-$, this implies a very efficient energy transfer from the above discussed vibrationally excited molecules to NO$^-$. 

What about more strongly bound anions? The boron anion, B$^-$, has a detachment energy of 0.28 eV, more than an order of magnitude larger than that of Ca$^-$. Here, the bending modes of H$_2$O and HCN do not suffice to ionize the anion and we resort to the antisymmetric stretch vibrations: 3755.8 cm$^{-1}$ and $A^A = 76$ s$^{-1}$ for H$_2$O, 3311.5 cm$^{-1}$ and $A^A = 74$ s$^{-1}$ for HCN, and 2359.15 cm$^{-1}$ and $A^A = 450$ s$^{-1}$ for CO$_2$ \cite{vib_Radzig_Smirnov}. At these energies the cross sections are $\sigma^B = 20, 25$ and $55$ Mb, respectively \cite{B_Anion_PD}. In spite of the larger frequencies and smaller cross sections, the energy transfer is still rather efficient. At 1 nm distance the vibrational ICD lifetimes are sub-nanosecond for water ($\tau = 0.86$ ns) and for hydrogen cyanide ($\tau = 0.43$ ns) and much shorter for carbon dioxide ($\tau = 0.81$ ps). Even at the truly large distance of 2 nm, the transfer is sub-nanosecond ($\tau = 0.52$ ns) for the latter molecule.

For even stronger bound anions, like the alkali Li$^-$ to Cs$^-$, two vibrational quanta are needed to enable the energy transfer. The binding energies of the alkali anions decrease smoothly from 0.618 eV for Li$^-$ to 0.4716 eV for Cs$^-$ \cite{Andersen_Review,EA_Schaefer}. We choose hydrogen fluoride (HF) as the vibrationally excited partner. The overtone $\nu=2$ amounts to 7750.8 cm$^{-1}$ and its coefficient is $A_{2,0}^A = 23$ s$^{-1}$ \cite{vib_Radzig_Smirnov}, and measurements of the absolute photodetachment cross sections are available and range for the respective energy from 130 Mb for Li$^-$ to 200 Mb for Cs$^-$ \cite{alkali_PD}. The resulting vibrational ICD lifetimes are similar for all alkali negative ions slightly decreasing from 7.8 ns for Li$^-$ to 5.1 ns for Cs$^-$ at R=1 nm. The energy transfer can be made even more efficient if one resorts to the 10$\rightarrow$8 vibrational transition of HF. Here, the vibrational energy is 5219 cm$^{-1}$ and the Einstein coefficient much larger $A_{10,8}^A = 560$ $s^{-1}$  The cross sections $\sigma^B$ are now smaller for the first alkali, 50 Mb for Li$^-$, but grow to 290 Mb for Cs$^-$ \cite{alkali_PD}. The energy transfer times now range from 0.17 ns for Li$^-$ to the short time of 30 ps for Cs$^-$. Even at R=2 nm the latter time is still fast (1.9 ns).

For Eq.(\ref{eq::7}) to be reliable, the distance R between the vibrating molecule and the anion should be much larger than the mean radius of the isolated negative ion. For an atomic anion this mean radius can be estimated from its binding energy \cite{anion_radius}. This mean radius is 0.62 nm for Ca$^-$, drops down to 0.18 nm for B$^-$ and further down to 0.14-0.12 nm for the alkalies, i.e., for Ca$^-$ at 1 nm the multipole expansion provides only a crude estimate and is an acceptable estimate at 2 nm while for B$^-$, Cs$^-$, Rb$^-$, K$^-$, Na$^-$ and Li$^-$ the multipole expansion can be expected to be very reliable at the distances applied. What to expect at shorter distances like in anionic clusters, for instance? For purely electronic ICD processes the lifetime computed \textit{ab initio} is usually even shorter than that predicted by the multipole expansion (see, e.g., \cite{Averbukh_Review}), and there is reason to believe that this trend also holds for vibrational ICD. However, one has to await further studies to find out what happens in real systems in nature.


It has been explicitly demonstrated that the energy transfer from vibrationally excited molecules to ionize a neighboring anion can be extremely efficient. The lifetime of the isolated vibrationally excited molecules is typically seconds to milliseconds and due to vibrational ICD it can decrease to nanoseconds and picoseconds and in favorable cases even to the femtosecond regime, and all of that at internuclear distances much beyond the distances at which the partners can have a bond. Scenarios for measurements could be clusters \cite{anion_Clusters1,anion_Clusters2} and cold merged beams \cite{cold_Ed_1}. By measuring the distribution of the emitted electrons one can discern between different molecules and neighbors. Finally, we mention that, in principle, the neighbor can also be an electronically excited {\it neutral} system whose lifetime is longer than the ICD time and which can be ionized by the vibrational energy transfer.

\section*{Acknowledgements} 
The author thanks K. Gokhberg, T. Jahnke and A. Kuleff for valuable contributions. Financial support by the DFG (research unit 1789) and by the European Research Council (ERC) (Advanced Investigator Grant No. 692657) is gratefully acknowledged



\end{document}